# S&P 500 returns revisited

Ivan O. Kitov, Oleg I. Kitov

**Abstract**
The predictions of the S&P 500 returns made in 2007 have been tested and the underlying models amended. The period between 2003 and 2008 should be described by the dependence of the S&P 500 stock market index on real GDP because the population pyramid was highly inaccurate. The 2008 trough and 2009 rally are well predicted by the original model, however. The rally will end in March/April 2010 and the S&P 500 level will be decreasing into 2011. This prediction should validate the model.

JEL Classification: G1, D4, J1,

**Introduction**

In this paper, we test and amend our 2007 model [1] the S&P 500 stock market index as a function of a specific age population, and thus, of real GDP per capita in the United States. Our approach is based on the assumption that stock exchange represents a gauge measuring the future states of real economy. The term "gauge" is used in its technical meaning in order to stress that the link between some aggregate measure of the stock market and real economic growth is a mechanical one. In this regard, the link does not depend on qualities of agents populating the economy but solely on their quantities. Then, the S&P 500 index covering ~75% of U.S. equities is a good approximate measure to describe the evolution of the economy as a whole. The accuracy of such a measure depends on the understanding of the forces behind real economic growth and the uncertainty associated with relevant measurements. Thus, when the trajectory of real (and nominal) economic growth is exactly known one can also predict aggregate stock market indices.

An accurate prediction of real economic growth resolves two major problems – the stock market obtains a tool for estimation of stock prices and, reciprocally, the uncertainty in GDP measurements might also be reduced. The discrepancy between measured GDP and that obtained from stock market indices should converge or, equivalently, the macroeconomic state of a developed economy should be exactly described by the aggregate stock market indices. The term "exactly" implies that increasing accuracy in the prediction of GDP unambiguously leads to the vanishing discrepancy between predicted and observed (aggregated) stock market indices. Both measures should converge to one curve.

Previously, we have found and validated a link between the change rate of a specific age population and real economic growth [2, 3]. Real GDP in the USA has three principal components of growth. First of these components is associated with the extensive growth in working age population and workforce. This component has been consistently positive in the USA and its input reaches almost 1 percentage point per year. When comparing the USA to other developed countries with low population growth, one should correct the USA figures by this extensive factor. Effectively, this is equivalent to the comparison of real GDP per capita values, GDPpc.

The intensive component of real growth, GDPpc, is driven by two sources. We have found that in the long-run developed economies are characterized by constant annual increment of real GDP per capita [4]. Statistically, the increment time series have no trend over the past 55 years of measurements. When averaged, the annual increments are close in leading economies. In terms of growth rate, asymptotic behavior in all developed economies is also similar – the rate is inversely proportional to the attained level of real GDP per capita. Therefore, the growth rate in all economies will be asymptotically approaching zero, as represented by the following relationship:

$$dln(GDPpc) = G_0/GDPpc \qquad (1)$$

, where $G_0$ is a country specific constant.

Third source of real GDP growth is responsible for short-term fluctuations around the trend defined by (1). This source is related to the change in a specific age population. In the United States this age is nine years as well as in the UK [5]. European countries and Japan are characterized by the age of eighteen years. The presence of the third source was validated using observations in the USA, the UK, Japan, France, Germany, New Zealand, and Austria. This list includes the biggest economics and also relatively small economies. To this end, the evolution of real GDP in the USA from 1960 to 2005 is well predicted by three principal variables - working age population (WAP), constant increment and the growth rate of nine-year-olds.

The working age population and the long-term trend are slowly evolving variables. Between 1985 and 2005, the mean growth rate of WAP in the USA was 1.19% with the largest value of 1.5% in 2000 and the smallest value of 0.8% in 1989. Both extremes might be biased by population revisions conducted by the Bureau of Labor Statistics after the censuses conducted in 1990 and 2000. The long-term trend in the USA, as defined by the constant annual increment, fell from 1.7 % in 1985 to 1.2% in 2005 [4]. In the first approximation, one can neglect the observed secular changes in both variables as small-amplitude ones, and treat them as constants. However, it is possible to extend the analysis and include actual behavior of both variables.

Having tight relations between stock market indices and economic growth and between economic growth and specific age population one can test all effects of the change in the specific age population on the stock market. The intuition behind this link is very simple - the population change induces positive or negative economic growth, which, in turn, is reflected in aggregate stock prices. All other financial, economic, demographic and social factors are neglected.

1. **Prediction of the S&P 500 returns between 1985 and 2003**

Under our framework, aggregate stock indices depend on real economic growth, and thus, on the population of a country-specific age. Before modeling the S&P 500 returns we would like to inspect raw data and discuss important features of the population age distribution. The US Census Bureau has been reporting monthly estimates of single year of age populations since March 1990. Before 1990, only quarterly and annual estimates are available. The methodology of estimation, as described by the Census Bureau [6], includes monthly statistics of net births and deaths. Basic period for these estimates is one quarter, however. All migration processes are described at an annual basis and then evenly distributed over months. The absence of accurate monthly population estimates leads to a higher uncertainty in the modeling. High autocorrelation is induced by smoothing and/or redistribution of the true changes over calendar quarters and age cohorts. Also, decennial censuses are used to revise the evolution of the population age pyramid across calendar years and age groups. These revisions additionally redistribute the counts in five- to ten-year-wide age groups in a way to equalize populations in neighboring cohorts, but introduce artificial steps between the groups. This procedure adds to the deterioration of the population estimates consistency as compared to the true distribution.

Our aim is to describe the S&P 500 returns using monthly estimates of the nine-year-old population in the USA. Due to the problems with the overall consistency the monthly population estimates were smoothed over neighboring calendar months and over the same months of adjacent calendar years. The former approach reduces large fluctuations associated with the death statistics for nine-year-olds. The latter one was inspired by the methodology of the US Census Bureau which revises monthly and quarterly population estimates in various age groups. It is worth noting that the revised population estimates are distorted by the errors of closure, which are different in various age groups in absolute and relative terms. As a result, the best representation of the monthly population estimates may vary between the age groups.

There is no doubt that high-frequency fluctuations of the stock market are driven by a multitude of factors including news issued by political and financial authorities, changes in weather conditions, and individual actions of stock market participants. These short-term fluctuations have negligible effect on the long-term evolution, however. The latter is more likely to be governed by macroeconomic variables.

In this study, the S&P 500 returns are represented by a running sum of monthly returns over twelve consecutive months. The monthly returns are calculated from the closing levels of the S&P 500 index. A natural time step is one month. The summation allows obtaining a smoother curve than that provided by the annual S&P 500 returns with one month step. Figure 1 demonstrates the difference between these two definitions.

As discussed in [7], the choice of closing levels for the calculation of returns might be not the best one. Due to inevitable fluctuations at daily and shorter time intervals the largest monthly return is always different from that represented by the closing levels. Figure 2 sheds some light on the degree of this uncertainty. In the left panel, the difference between the monthly mean and closing levels of S&P 500 is shown, as normalized to the mean level for given month. The normalized difference is relatively stable over the period between 1980 and 2009 with several outbursts associated with the sessions of very high volatility. The average difference is ~0.003 with standard deviation of 0.02. Because of the discrepancy, the return defined by the monthly means differs from that obtained from the closing levels, as best demonstrates the right panel in Figure 2, where cumulative returns are shown. In the long run, the cumulative curves diverge with the "mean" curve consistently below the "closing" one.

The inter-month fluctuations are also substantial. In Figure 3, monthly standard deviations are depicted for the daily closing levels of S&P 500. There are two months with extremely large standard deviations: October 1987 and October 2008. Otherwise, the deviations reside between 0 and 0.04 with the mean value of 0.02. The bigger is the inter-month fluctuation the larger is the return (positive or negative).

Having introduced this measure of the S&P returns and the driving force we now formulate relevant mechanical link between them. The annual S&P 500 returns are modeled using the monthly growth rates of the nine-year-old population estimates. For example, the annual S&P 500 return for June 1995, i.e. the sum of monthly returns between July 1994 and June 1995, is proportional to the ratio of monthly population estimates for May and June 1995. The population estimates are smoothed according to the following procedure. The trial-and-error method demonstrated that the best monthly estimates of the number of nine-year-olds for our purposes are those averaged over five adjacent years of age, with the nine-years-olds in the center. For example, the number of nine-year-olds in June 1995, $N_9(1995.5)$ is estimated as the average value of 7-, 8-, 9-, 10-, and 11-year-olds in June 1995:

$N_9(1995.5)= [N_7(1995.5)+N_8(1995.5)+N_9(1995.5)+N_{10}(1995.5)+N_{11}(1995.5)]/5.$

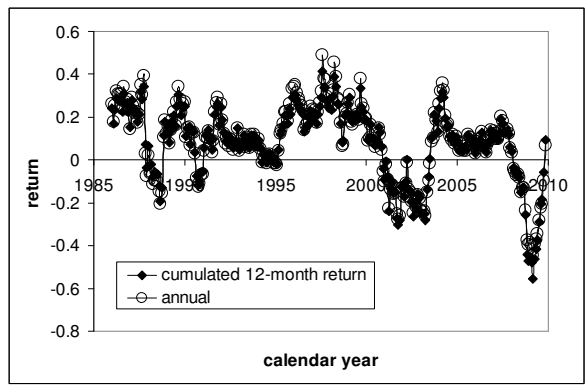

Figure 1. Comparison of the annual S&P 500 returns and that cumulated during the previous twelve months as a sum of monthly returns. The annual curve is of a higher volatility.

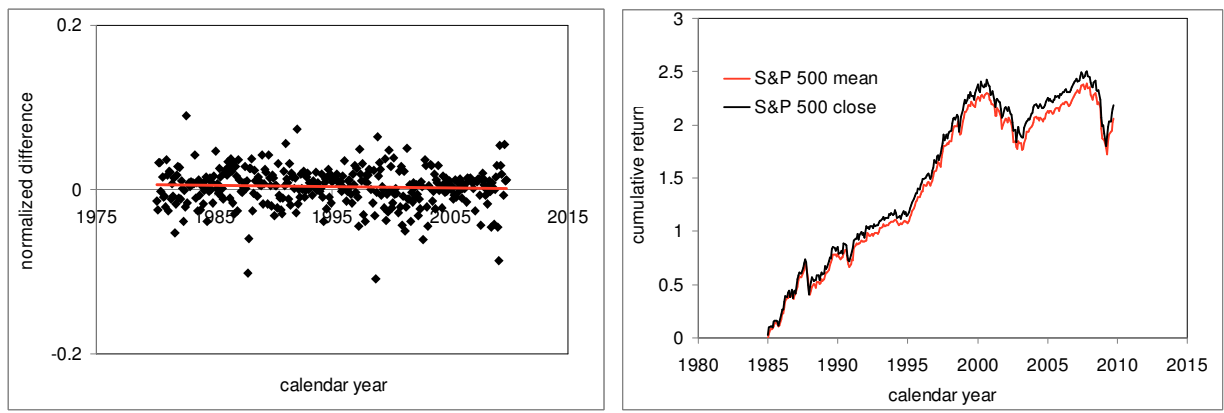

Figure 2. *Left panel*. The difference between the monthly mean and closing level of S&P 500 normalized to the mean level for given month. The normalized difference is relatively stable over the period between 1980 and 2009 with several outbursts associated with rare sessions of high volatility. *Right panel*. The difference between cumulative returns as defined by the monthly mean and closing S&P 500 index.

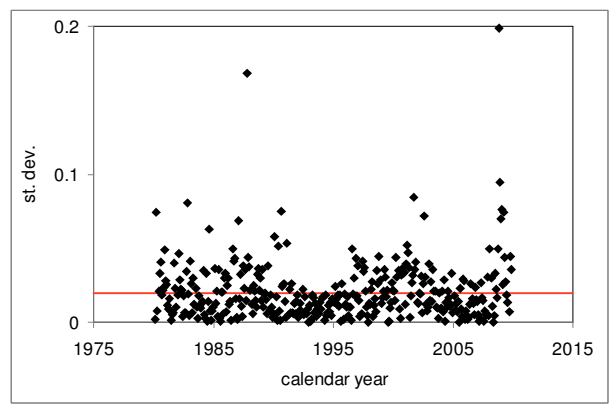

Figure 3. Monthly standard deviations of the daily closing levels of S&P 500. There are two months with extremely large standard deviation (volatility): October 1987 and October 2008. Mean value of the standard deviations over the period from 1980 to 2009 is 0.02.

The intuition behind this representation is associated with the balancing of single-year-of-age populations in five- to ten-year-wide age groups, as carried out by the US Census Bureau. Effectively, these revisions to the single-year populations result in the trade-off between monthly estimates for adjacent ages. The approach used in the study recovers a bigger part of the true monthly values, but this problem has to be investigated in detail.

The introduced approximation of the number of nine-year-olds, $N_9(t)$, is justified also by the excellent prediction of the S&P 500 returns for the period where the monthly estimates are available, except the years after 2003. The relationship linking the S&P 500 returns and the change rate of the number of nine-year-olds is as follows:

$$R_p(t) = v_1 dlnN_9(t) + v_2 \qquad (2)$$

, where $R_p(t)$ is the predicted return; coefficients $v_1$ and $v_2$ are determined empirically.

There are two different time series for the number of nine-year-olds between 1990 and 2000: *post*censal and *inter*censal. The former is obtained by the "inflation-deflation" component method using contemporary estimates of total deaths and net migration including the net movement of the US Armed Forces overseas. The start point for the single-year-of- age time series is the counts in the 1990 Census.

The intercensal time series is estimated using the population counts in the 1990 and 2000 Censuses as the start and end points. The single-year-of-age intercensal time series are obtained from corresponding postcensal series by a proportional redistribution of the errors of closure over the ten years between the censuses on a daily basis. Since the errors of closure are age dependent, adjacent time series may converge or diverge by several per cent. For example, the postcensal estimate of the population under five years of age for April 2000 was underestimated by about 1% relative to the census count. The population between 5 and 13 years of age was underestimated by 3.5% and that between 14 and 17 years of age was underestimated by 2.4%. An additional disturbance to the monthly estimates of the age pyramid is introduced by the adjustment of the sum of the single year estimates to the total population obtained by an independent procedure.

After April 2000, only the postcensal population estimates are available. There are several vintages of these estimates available for previous years, however, which use most recent information on the past estimates of the rate of deaths and migration. Therefore, no postcensal estimate is final and further revisions are likely for all monthly estimates.

When a single year of age population is used for the prediction of S&P 500 returns, the difference between the post- and intercensal populations is expressed only in a synchronized and proportional change in level. The difference in the change rate due to the error of closure is evenly distributed over months and practically is not visible in the monthly estimates. There is a several percent difference for the cumulative curves as dictated by the error of closure for the nine-year-olds.

As discussed above, the (postcensal) monthly estimates of nine-year-olds in our study are obtained using also the numbers of 7- through 11-year-olds. All these ages are inside the US Census Bureau specified age group between 5 and 13 years. Using these modified monthly estimates of $N_9(t)$ one can predict the annual S&P 500 returns according to (2). Figure 4 displays the observed and predicted time series for the S&P 500 returns between 1990 and 2003. The latter is obtained by varying coefficient $v_1$ and $v_2$ to minimize the RMS difference between two time series. No formal minimization procedure was used, however, and, for the postcensal population estimates, the best manually obtained coefficients $v_1=170$ and $v_2=-0.04$ provide the mean of -0.0003 with standard error of 0.082. Visually, the predicted and measured curves are similar. In the long run, high-frequency fluctuations in both series are cumulated to zero. The average S&P 500 return (according to the definition accepted in our study) for the same period is 0.16 with standard deviation of 0.10. The predicted time series is characterized by the average value of 0.158 and standard deviation of 0.091.

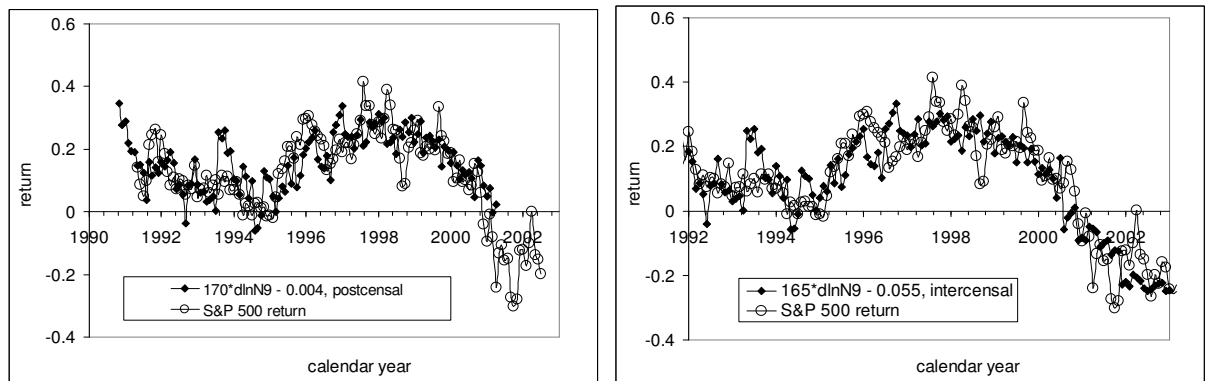

Figure 4. *Left panel*: Comparison of the observed and predicted S&P 500 returns between 1990 and 2000. The latter is obtained using the postcensal estimates of the 9-year-olds. RMS difference between the curves for the period between 1991 and 2001 is 0.082 with mean value only -0.0003. *Right panel*: Comparison of observed and predicted S&P 500 returns between 1992 and 2003. The latter is obtained using the intercensal estimates of the 9-year-olds.

Right panel in Figure 4 displays the predicted curve obtained using the intercensal estimates of nine-year-olds between 1992 and 2003. Three years between 2000 and 2003 are obviously postcensal estimates, but they are of a higher accuracy due to their closeness to the single year of age counts in the 2000 census. The best fit coefficients $v_1=165$ and $v_2=-0.055$ provide standard error of 0.085.

The post- and intercensal estimates of $N_9$ provide a consistent description of the S&P 500 returns between 1991 and 2003. There is a good opportunity, however, to obtain a relatively accurate prediction of the returns at time horizons between one and nine years using population estimates for younger ages as a proxy to the number of nine-year-olds. For example, the number of seven-year-olds provides a good approximation to the monthly increments of the number of nine-year-olds, $dN_9$. Both increments are shown in Figure 5. The largest difference between the curves is observed around 2000, when the Census Bureau revised all population estimates. Previously, we approximated $N_9$ by averaging of five consecutive cohorts between seven and eleven years of age. Hence, the estimate of $N_7$ includes the ages from five to nine.

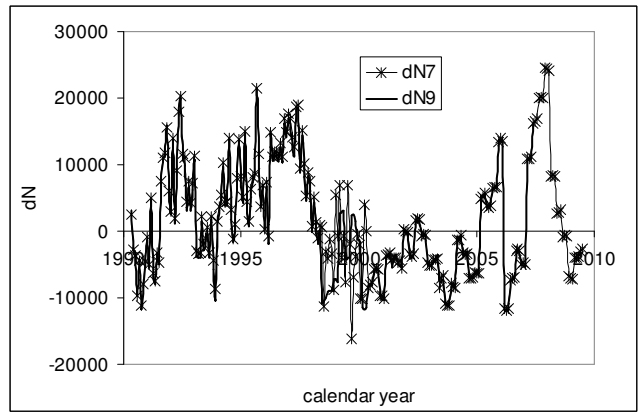

Figure 5. Comparison of monthly increments of the number of 9-year-olds, $dN_9$, and of the number of 7-year-olds, $dN_7$, shifted two years ahead.

Figure 6 displays the measured and predicted from $dN_7$ curves for the S&P 500 returns. The latter is a forecast at a two-year horizon. The best-fit coefficients are $v_1$=165 and $v_2$=-0.06. The period between 2001 and 2003 is described less accurately as expected from the difference between $dN_9$ and $dN_7$ around 2000. As a result, the model residual for the period from 1992 to 2003 has a higher standard deviation of 0.088. With the current population estimates it is possible to extend the forecasting horizon to nine years, with a slightly degrading accuracy. Population projections allow obtaining even longer predictions for the S&P 500 returns.

At a two-year horizon, the measured S&P 500 return is characterized by standard deviation of 0.18, i.e. twice as large as in the predicted time series. Effectively, these values demonstrate the predictability of the S&P 500 returns that is helpful for all stock market participants.

Having the monthly population estimates after April 1990 it is possible to extrapolate the prediction of the S&P 500 returns back in the past using older populations. The intuition behind this approach is the same as for the prediction of the future returns using younger ages - the monthly increments change slowly for the population with a fixed year of birth.

First, the averaging over five adjacent ages has been tested for the monthly estimates of older ages. Unfortunately, it gave results inferior to those for the period between 1990 and 2003. The deterioration of the predictive power is likely associated with the revision procedures applied by the Census Bureau to the population estimates in older age groups between 14 and 17 years and between 18 and 24 years. For the extrapolation of the 9-year-olds time series by four and more years in the past, the youngest of two or both age groups have to be used in the five-year averaging intervals.

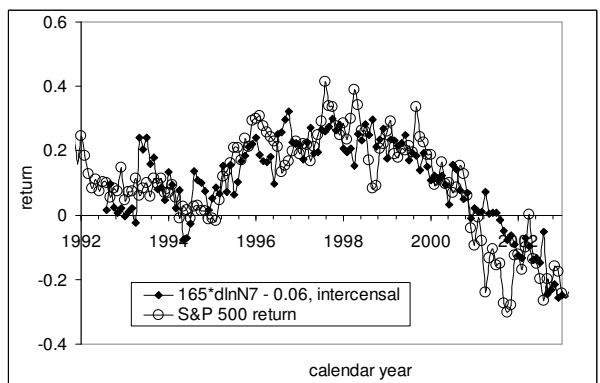

Figure 6. Comparison of the observed and predicted S&P 500 returns between 1992 and 2003. The latter is obtained using the intercensal estimate of 7-year-olds with $v_1$=165 and $v_2$=-0.06.

We have tried several alternative smoothing techniques. The best one is based on the averaging of neighboring months for the same age. When twelve successive moths are used, this approach is identical to the cumulative S&P 500 return for the previous 12 months. For shorter averaging windows the result is very similar to that for 12 months, as Figure 7 demonstrates. As an example, we used MA(4) and the number of seventeen-year-olds to predict the S&P 500 returns before 1991.

Figure 8 depicts the measured and predicted S&P 500 return between 1984 and 1995. The latter is shifted by eight years back relative to its natural position. The best-fit coefficients $v_1$=35 and $v_2$=0.089 are different from those obtained from five consecutive years. Despite the presence of high-frequency fluctuations the predicted curve repeats the most prominent features of the measured one. Notice an almost precise prediction of time and amplitude of the stock market crash in 1987. The difference between the measured and predicted curves is presented in the right panel of Figure 8. Standard error for the period from 1985 to 1993 is only 0.093, which is much better than that obtained using a naïve (random walk) prediction at an eight-year horizon.

The monthly estimates of S&P 500 returns provide a dynamic view, which is characterized by high-frequency fluctuations not related to the long-term equilibrium link between the stock market index and the number of 9-year-olds. Our model guarantees that the residual errors are canceled out in the long run. Therefore, relevant cumulative returns should have a strong tendency to converge and demonstrate the unbiased long-term relationship between the measured and observed S&P 500 returns. Figure 9 displays the observed cumulative curve and that predicted from $N_9$ and $N_{17}$, as described above. The curves are characterized by a few small-amplitude deviations which are compensated at shorter time intervals. Otherwise, two curves coincide, i.e. the long-term evolution of the stock market can be replaced by the evolution of the number of nine-year-olds.

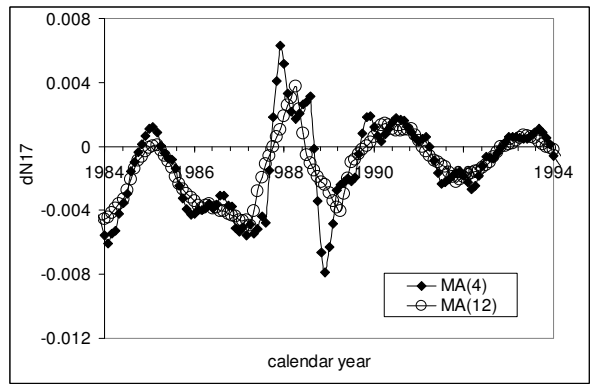

Figure 7. Comparison of MA(4) and MA(12) as applied to the number of 17-year-olds.

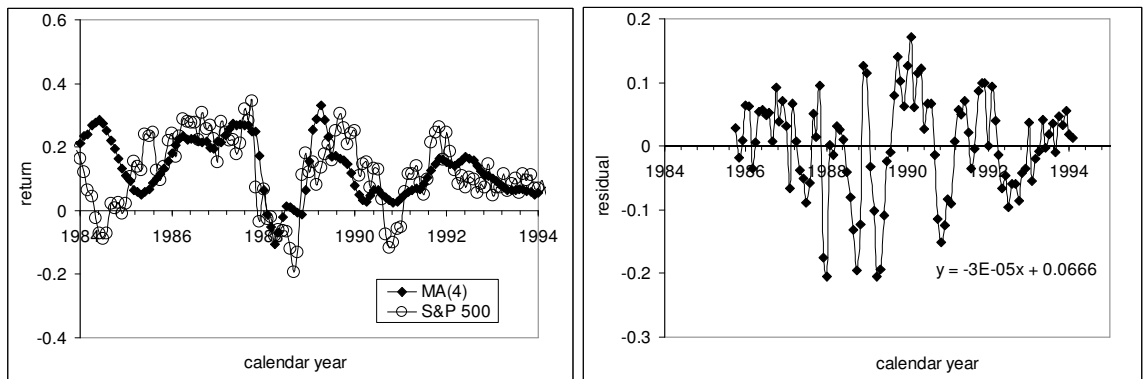

Figure 8. *Left panel*: Comparison of the observed and predicted S&P 500 returns between 1985 and 1994. The latter is obtained using the intercensal estimate of 17-year-olds. *Right panel*: The difference between the measured and predicted S&P 500 returns for the period between July 1985 and February 1994.

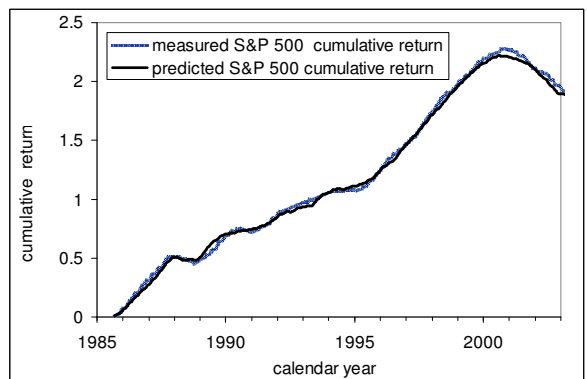

Figure 9. Comparison of the measured and predicted cumulative S&P 500 returns as obtained from $N_9$ and $N_{17}$.

The most prominent features, such as periods of near-zero and negative returns, are well described. The periods of rapid growth are also predicted. All these features are modeled using only one parameter - the number of 9-year-olds. This is an ultimately parsimonious model, which also provides accurate forecasts at various time horizons. The period of such an excellent description finished in April 2003, however, after a comprehensive revision to population estimates of the 2000 census. We expect that the next comprehensive revision in 2013 (after the 2010 census) will provide more accurate population estimates for the period between 2003 and 2013. While these accurate estimates are not available we replace $N_9$ with GDPpc, as discussed in Section 3.

## 2. Cointegration test

We have revealed a high degree of similarity between the observed, $R_o(t)$, and predicted, $R_p(t)$, S&P 500 returns, both dynamic and cumulative. Formal econometric tests may additionally validate the link between the stock market index and the number of nine-year-olds. In this Section, we test the existence of a long-term equilibrium (cointegration) relation between the measured and predicted S&P 500 returns during the period from 1985 to 2003.

According to Granger and Newbold [8], the technique of linear regression for obtaining statistical estimates and inferences related to time series is applicable only to stationary time series. Two or more nonstationary series can be regressed only in the case when there exists a cointegration relation between them [9].

There are four time series to be tested for unit roots - the measured and predicted S&P 500 returns and their first differences. The Augmented Dickey-Fuller (ADF) and the modified DF t-test using a generalized least-squares regression (DF-GLS) are used. These tests should provide adequate results for the series consisting of 207 monthly readings.

Results of unit root tests for these four series are listed in Tables 1 and 2. Both original series are characterized by the presence of unit roots - the test values are significantly larger than the 1% critical values. Both first differences have no unit roots and thus are stationary. In the ADF tests, trend specification is constant and the maximum lag order is 3. In the DF-GLS tests, the maximum lag is 4 and the same trend specification is used.

The presence of unit roots in the original series and their absence in the first differences evidences that the former series are integrated of order 1, I(1). This fact implies that cointegration tests have to be carried out. Otherwise, regression is potentially a spurious one.

An assumption that the measured and predicted (i.e. the change rate of nine-year-olds) returns are two cointegrated non-stationary time series is equivalent to the assumption that their difference, $\varepsilon(t) = R_o(t) - R_p(t)$, is a stationary or I(0) process. Therefore, it is natural to test the difference for unit roots. If $\varepsilon(t)$ is a non-stationary variable having a unit root, the null of the existence of a cointegrating relation can be rejected. Such a test is associated with the Engle-Granger approach [10], which requires $R_o(t)$ to be regressed on $R_p(t)$, as the first step. It is worth noting that the predicted variable is obtained by a procedure similar to that of linear regression and provides the best visual fit between corresponding curves. So, we skip the regression step.

Table 1. Results of unit root tests for the original time series – measured and predicted S&P 500 returns. Both series are characterized by the presence of unit roots.

| Test | Lag | Time series | | 1% critical |
|---|---|---|---|---|
| | | predicted | measured | |
| ADF | 0 | -2.65 | -2.07 | -3.47 |
| | 1 | -2.58 | -1.63 | -3.47 |
| DF-GLS | 1 | -2.69 | -2.47 | -3.48 |
| | 2 | -2.34 | -1.99 | -3.48 |

Table 2. Results of unit root tests for the first differences of the original time series – measured and predicted S&P 500 returns. Both series are integrated of order 0, I(0).

| Test | Lag | Time series | | 1% critical |
|---|---|---|---|---|
| | | predicted | measured | |
| ADF | 0 | -15.6* | -16.6* | -3.47 |
| | 1 | -12.0* | -9.2* | -3.47 |
| | 2 | -9.2* | -7.8* | -3.47 |
| | 3 | -8.7* | -7.7* | -3.47 |
| DF-GLS | 1 | -7.55* | -9.1* | -3.48 |
| | 2 | -7.38* | -7.6* | -3.48 |
| | 3 | -7.33* | -7.5* | -3.48 |
| | 4 | -6.2* | -7.7* | -3.48 |

The Engle-Granger approach is most reliable and effective when one of the two involved variables is weakly exogenous, i.e. is driven by some forces not associated with the second variable. This is the case for the S&P 500 returns and the number of 9-year-olds. The latter variable is hardly to be driven by the former one.

The results of the ADF and DF-GLS tests listed in Table 3 indicate the absence of unit roots in the difference between the measured and predicted series. Since the predicted series are constructed in the assumption of a zero average difference, trend specification in the tests is *none*, and the maximum lag order is 3. The units root tests give strong evidences in favor of the existence of a cointegrating relation between the measured and predicted time series. From an econometric point of view, it is difficult to deny that the number of 9-year-olds is the only defining factor behind the observed long-term behavior of S&P 500.

The Johansen [11] approach is based on the maximum likelihood estimation procedure and tests for the number of cointegrating relations in the vector-autoregressive representation. The Johansen approach allows simultaneous testing for the existence of cointegrating relations and determining their number (rank). For two variables, only one cointegrating relation is possible. When cointegration rank is 0, any linear combination of the two variables is a non-stationary process. When rank is 2, both variables have to be stationary. When the Johansen test results in rank 1, a cointegrating relation between involved variables does exist.

Table 3. Results of unit root tests of the differences between the predicted and observed time series. There is no unit root in the difference.

| Test | Lag | Difference | 1% critical |
|---|---|---|---|
| ADF | 0 | -7.6* | -3.47 |
|  | 1 | -6.8* | -3.47 |
|  | 2 | -6.4* | -3.47 |
|  | 3 | -6.8* | -3.48 |
| DF-GLS | 1 | -6.7* | -3.48 |
|  | 2 | -6.3* | -3.48 |
|  | 3 | -6.7* | -3.48 |

In the Johansen approach, one has first to analyze specific properties of the underlying vector auto-regression (VAR) model for the two variables. Table 4 lists selection statistics for the pre-estimated maximum lag order in the VAR. Standard trace statistics is extended by several information criteria: the final prediction error, FPE, the Akaike information criterion, AIC, the Schwarz Bayesian information criterion - SBIC, and the Hannan and Quinn information criterion, HQIC. All tests and information criteria indicate the maximum pre-estimated lag order 3 for the VARs and vector error-correction models (VECM). Therefore, maximum lag order 3 was used in the Johansen tests along with *constant* as a trend specification.

Table 4. Pre-estimation lag order selection statistics. All tests and information criteria indicate the maximum lag order 3 as an optimal one for the VARs and VECMs.

| Lag | LR | FPE | AIC | HQIC | SBIC |
|---|---|---|---|---|---|
| 0 |  | 0.0065 | -2.19 | -2.18 | -2.16 |
| 1 | 211 | 0.0023 | -3.22 | -3.20 | -3.17 |
| 2 | 1.39 | 0.0023 | -3.21 | -3.19 | -3.15 |
| 3 | 9.5* | 0.0022* | -3.254* | -3.22* | -3.17* |
| 4 | 1.7 | 0.0022 | -3.253 | -3.21 | -3.15 |

Table 5. Johansen test for cointegration rank for the measure and predicted time series. Maximum lag order is 3.

| Trend specification | Rank | Eigenvalue | SBIC | HQIC | Trace statistics | 5% critical value |
|---|---|---|---|---|---|---|
| none | 1 | 0.196 | -5.79* | -5.93* | 2.93* | 3.84 |
| rconstant | 1 | 0.196 | -5.65* | -5.92* | 3.24* | 9.25 |
| constant | 1 | 0.196 | -5.74* | -5.90* | 2.67* | 3.76 |

Table 5 lists results of the Johansen tests – in both cases cointegrating rank is 1, i.e. there exists a long-term equilibrium relation between the observed and predicted S&P 500 returns. We do not test for causality direction between these variables because the only possible way of influence, if it exists, is absolutely obvious.

The measured and predicted time series are cointegrated. Therefore, the estimates of the goodness-of-fit and RMSE in various statistical representations have to be valid and provide important information on the accuracy of corresponding measurements and the relation itself. The VAR representation is characterized by $R^2$=0.89 and RMSE=0.047, with mean annual return of 0.18 for the same period. The standard error of ~5 percentage points has been reached due to strong noise suppression. In practice, the AR is a version of a weighted moving average, which optimizes noise suppression throughout the whole series. Simple linear regression provides a lower $R^2$=0.66 and a larger RMSE = 0.07.

### 3. S&P 500 returns and real GDP

In Section 2, we failed to predict the S&P 500 return beyond 2003. The failure might be associated with a structural break according to two different mechanisms. First, the break is induced by some real economic processes, i.e. may reflect the change in the inherent link between true values of the studied variables – the number of 9-year-olds and S&P 500 returns. This new link is likely to be linear, as was the link observed before 2003. Second mechanism is related to some changes in the population measuring procedure. In this case, the structural break is artificial and the relationship for the period before 2003 can be easily transported in a scaled version to the period after 2003. Is there a possibility to distinguish between these two mechanisms?

As discussed above, there exists a trade-off between the growth rate of real GDP pre capita and the change rate of the number of 9-year-olds. Corresponding relationship should work in both directions and the number of 9-year-olds can be estimated from GDP measurements. So, one can replace $N_9(t)$ with *GDPpc(t)* in (2), taking into account that second term in the relationship between real GDP per capita and population is constant.

Figure 10 displays the observed S&P 500 returns and those obtained using real GDP, as presented by the US Bureau of Economic Analysis. The observed returns are presented by MA(12) of the monthly returns. The predicted returns are obtained from the following relationship:

*$R_p(t)$ = 0.62\*dln(GDPpc(t)) - 0.0094*,

where *GDPpc(t)* is represented by MA(6) of the (annualized) growth rate during or six previous months or two quarters as only quarterly readings of real GDP are available. (According to Figure 10, the level of GDP between 2000 and 2003 was underestimated.)

The period after 1996 is relatively well predicted including the increase in 2003. Therefore, it is reasonable to assume that the 9-year-old population was not well estimated by the US Census Bureau after 2003. This conclusion is supported by the cointegration test conducted for real GDP per capita and the charge rate of the number of 9-year-olds, which proves the existence of a long-term equilibrium linear relation between these two variables since the early 1960s [2]. As a result, one can use either $N_9(t)$ or $GDPpc(t)$ for modeling of the S&P 500 returns, where appropriate. Obviously, the $GDPpc(t)$ is consistent with the S&P 500 returns after 2003.

There is a concern related to the accuracy of population and real GDP measurement in 2006. In Figure 10, the predicted curve fell to -0.075 in the third quarter of 2006. There was no significant decrease in the S&P 500 returns during the same period. A possible reason for the discrepancy is that the real GDP was underestimated. This issue should be resolved in the next comprehensive revision to the GDP.

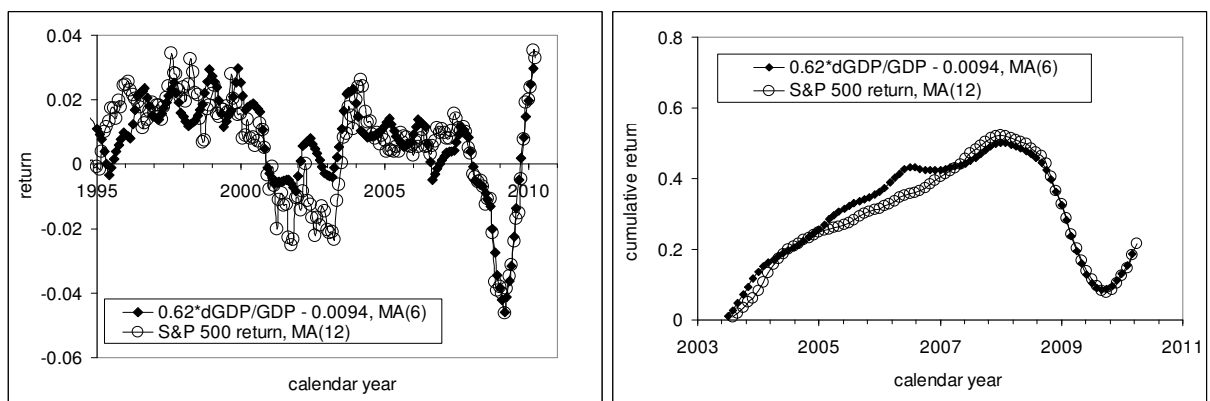

Figure 10. The observed and predicted S&P 500 returns. The latter are obtained using quarterly readings of the growth rate of real GDP. One may expect rapid economic growth in 2010.

A striking feature in Figure 10 is the agreement between the annual curves in 2008 and 2009. The GDP readings predict the S&P 500 returns in time and amplitude. Moreover, the S&P index leads the GDP curve and predicts a rapid real economic growth in 2010. This is a good prediction to validate the link. All in all, real GDP per capita is a good predictor of the S&P 500 returns, especially during periods of big changes.

### 4. Discussion

The number of nine-year-olds, in one form or another, demonstrates the predictive power far beyond that of the naïve model. Statistically, the S&P stock market index is not an unpredictable one. Further improvements are possible through the increasing accuracy of population measurements and the use of advanced statistical methods of noise suppression.

Figure 11 summarizes three best fit models for the following segments: 1985-1992, 1992-2002, and 2002-2009. The cumulative curves coincide over the whole period with the small deviations likely related to measurement noise. The predictability of the S&P 500 returns follows from the inherent properties of the predictor – the population estimates of the number of nine-year-olds can be accurately extrapolated by younger cohorts. In essence, the predicted curve leads the observed one by nine(!) years.

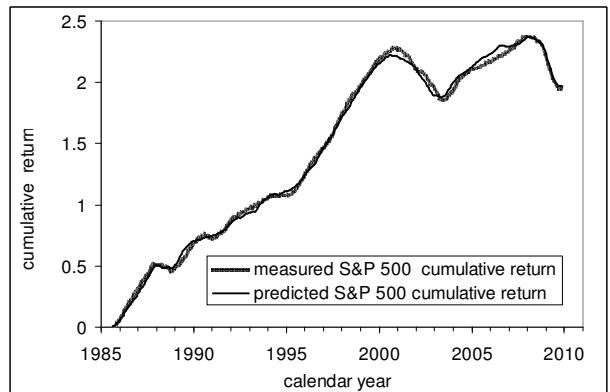

Figure 11. The observed and predicted cumulative S&P 500 return from 1985 to 2009.

We have to admit that the period between 2003 and 2008 is not well described when the monthly population estimates are used. The measured S&P 500 return started to grow in April 2003. By February 2004 it increased by 0.61 – from -0.28 to +0.33. There is no sign of such an increase in the nine-year-old postcensal estimates. Younger ages, some of them explicitly counted during the 2000 census, also do not demonstrate any significant steps in 2003.

At the same time the number of three-year-olds, $N_3$, as reported by the Census Bureau, might be a useful predictor for the period after 2008. Figure 12 displays the observed S&P 500 return and that predicted according the following relationship:

$$R_p(t) = 160 d\ln N_3(t) - 0.23$$

where $N_3(t)$ is taken straight from the Census Bureau tables without additional smoothing or correction, but the predicted series is smoothed with MA(6). Notice that $v_1$=160, i.e. very close to that in the models obtained in Section 1.

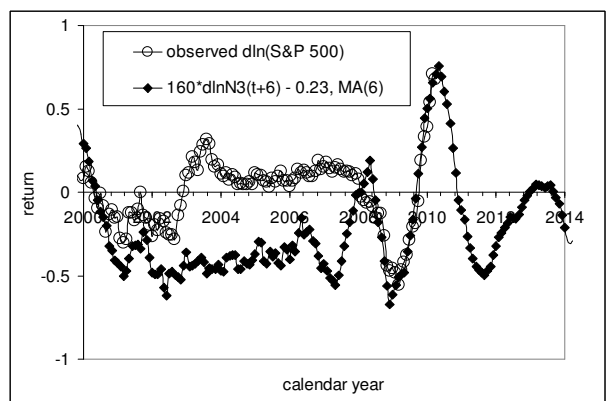

Figure 12. The observed S&P 500 return and that predicted from $N_3$ from 2000 to 2014. (The last measured value corresponds to March 2010.)

Before 2008, the prediction is poor. But the financial crisis and recession are well predicted in time and amplitude. The last reading in Figure 12 is for December 2009. The number of three-year-olds predicts the growth in S&P 500 to continue till March/April 2010 (we finished the paper in March 2010). When the peak is reached the (annual) returns will start to decline during the next 16 months to the level of -0.5. It is only in 2012 a new (weak) rally will start. This is also a prediction to validate the concept.

The Efficient Markets Hypothesis implies that current stock prices always 'fully reflect' available information. The only reason for a price to change is the arrival of 'news' or unanticipated events. Same hypothesis underlies the pricing models for goods, services and commodities. A common assumption in all mainstream models is that the price follows a geometric Brownian motion, and the stochastic evolution is equivalent to the unpredictability of prices.

In terms of physics a process may be stochastic but fully predictable. A banal example is thunder. One can not predict the time of a lightning discharge, but easily the arrival of the sound it generates. The arrival time depends only on the distance to the source. So, the time series of thunder sound arrivals to a given point is a stochastic one, but a fully predictable one. The trick consists in the presence of two independent physical variables - light and sound, which are tied by a causal relation, and corresponding channels of signal propagation with different speeds.

Our results demonstrate that the prices of goods and services, commodities and stocks are driven by external forces of different nature. We presume that the link between the forces and prices is a causal one, and the latter can be predicted when the former are measurable. Scientifically, any physical causality can only be expressed in statistical terms. A number of cointegration test and regressions showed a high level of confidence that supports our general conclusion - the revealed links are of a causal nature.

In addition to the obtained statistical estimates for the past observation all models predict prices in the future. This is an out-of-sample prediction, which will be used to validate the models. Therefore, we will trace, analyze and report the results of the model performance for all predictions. Relevant quantitative estimates will be updated when new data, revisions to old data and/or amended models are available.